\documentclass[peerreview]{IEEEtran}
\renewcommand{\baselinestretch}{1.5}
\usepackage{amsmath,amssymb}

\usepackage{amssymb}
\usepackage{bm}
\usepackage[dvipdfmx]{graphicx}
\usepackage{color}

\graphicspath{{figure/}}

\newcommand{\defeq}{\stackrel{\triangle}{=}}

\newtheorem{definition}{Definition}

\newtheorem{corollary}{Corollary}
\newtheorem{theorem}{Theorem}

\def\qed{\hfill $\Box$} 
\usepackage{setspace}

\begin{document}
\title{Bounds on Asymptotic Rate of Capacitive Crosstalk Avoidance Codes for On-chip Buses} 
\author{
  \IEEEauthorblockN{Tadashi Wadayama and Taisuke Izumi} \\
  \IEEEauthorblockA{Nagoya Institute of Technology,   Japan\\
      Email: wadayama@nitech.ac.jp, t-izumi@nitech.ac.jp} 
}

\maketitle
\begin{abstract}
In order to prevent the capacitive crosstalk in on-chip buses, several types of
{\em capacitive crosstalk avoidance codes}  have been devised.
These codes are designed to prohibit transition patterns prone to the capacity crosstalk 
from any consecutive two words transmitted to on-chip buses.
This paper provides a rigorous analysis on the asymptotic rate of $(p,q)$-transition free word sequences 
under the assumption that coding is based on 
a pair of a stateful encoder and a {\em stateless decoder}.
The symbols $p$ and $q$ represent $k$-bit transition patterns that should not 
be appeared in any consecutive two words at the same adjacent $k$-bit positions.
It is proved that the maximum rate of the sequences equals to the {\em subgraph domatic number} of
$(p,q)$-transition free graph. Based on the theoretical results on the subgraph domatic partition 
problem, 
a pair of lower and upper bounds on the asymptotic rate is derived.
We also present that the asymptotic rate $-2 + \log_2 \left(3 + \sqrt{17}  \right) \simeq 0.8325$ 
is achievable for the $p={\tt 01} \leftrightarrow q={\tt 10}$ transition free word sequences.
\end{abstract}

\section{Introduction}

A VLSI-chip consist of several components such as CPU cores and 
reliable interconnection between them are essential to build a robust system on-chip.
Inter components usually communicate with each other via an on-chip bus, which is a bundle of lines.
For example, a CPU chip with multiple cores equips data buses for 
exchanging data among  the cores.
In recent VLSI technology,  shrinking the circuit size is still of great importance 
because it leads to better yields, lower power consumption, and faster computation.
However, shrinkage of VLSI-chip tends to cause a negative impact on reliable inter-component communication.
In order to decrease the circuit size,  we have to make the line width and line spacing narrower.
It results in increased capacitance between adjacent lines in on-chip buses \cite{NoC}.
When the clock frequency is sufficiently high, the capacitive couplings between two adjacent lines
become nonnegligible. The capacitive coupling induces the {\em capacitive crosstalk}, which significantly degrades the reliability of
data exchange over buses. 

Assume that we have 3 adjacent lines $L_1, L_2, L_3$ in an on-chip bus.  The center line $L_2$ 
is called a {\em victim line}. Between $L_1$  and $L_2$ (and also $L_2$  and $L_3$), there are capacitive couplings.
Consider the situation where the sender component changes the signals emitted to the bus from 
$(L_1,L_2,L_3) = ({\tt 0,1,0})$ to $(L_1,L_2,L_3) = ({\tt 1,0,1})$ at a certain time instant.
The abrupt increase of the voltage in $L_1$ and $L_3$ induces transient current flows from the both side lines 
$L_1$ and $L_3$ to the
victim line $L_2$ through the capacitive coupling between them. As a result, 
the timing of the voltage transition in the victim line is delayed against others at the receiver component.
This phenomenon is called the {\em capacitive delay}, which is one of the major harmful effects of capacitive couplings.

In order to avoid or to weaken the effect of the capacitive crosstalk,  several techniques for avoiding the capacitive crosstalk 
have been devised. A simple method is to insert several ground lines into the buses to shield 
the signal lines. This method is easy to implement but it cannot provide an optimal solution in terms of space efficiency.
Another promising method for preventing the capacitive crosstalk is to exploit 
{\em capacitive crosstalk avoidance codes} \cite{NoC}. The main idea of capacitive crosstalk avoidance codes 
is to prohibit transition patterns prone to the capacity crosstalk 
from any consecutive two words transmitted to a bus. 
For example, if two consecutive words do not have any adjacent transition ${\tt 010} \leftrightarrow {\tt 101}$,
the immunity to the capacitive crosstalk is expected to be improved \cite{NoC}.

Pande et al. \cite{Pande},  Sridhara and Shanbhag \cite{SS2007} presented coding schemes 
satisfing the condition that 
a codeword having the pattern ${\tt 010}$ (resp. ${\tt 101}$) are not followed by 
a codeword having the pattern ${\tt 101}$ (resp. ${\tt 010}$) at the same bit positions.
They call the codes satisfying the above constraint {\em forbidden overlap codes (FOC)}.
Another type of a constraint is also discussed by the same authors.
The coding to avoid the transition patterns ${\tt 10} \leftrightarrow {\tt 10}$ is said to 
be forbidden transition coding (FTC) \cite{Pande} \cite{SS2007}.  
For example, Pande et al. \cite{Pande}
proposed a simple 3-bit input/4-bit output stateless FTC.
Recently, Nisen and Kudekar presented an advanced joint FTC and ECC \cite{Niesen} and showed
a density evolution analysis.

This paper provides a rigorous analysis on the asymptotic rate of the $(p,q)$-transition free word sequences 
under the assumption that coding is based on 
a pair of a stateful encoder and a  {\em stateless decoder}.
The $(p,q)$-transition free word sequences is a natural generalization of FOC and FTC.
The symbols $p$ and $q$ represent $k$-bit transition patterns that should not 
be appeared in any consecutive two words at the same adjacent $k$-bit positions.
The term ``asymptotic'' represents the situation where the word length grows to infinity.

Victor and Keutzer \cite{Victor} presented rate analyses for  the $({\tt 10, 01})$-transition free word sequences
in the case where both of an encoder and a decoder are stateful,  and in the case where both of them are stateless. 
The asymptotic rates for combinations of a stateful encoder and a stateless decoder  
remains to be studied and thus it brings us a theoretical interest and challenge. 
The stateless decoder has a significant practical advantage over the stateful decoder because
it can prevent error propagation at the decoder caused by decoding errors.
\section{Preliminaries}

The argument presented in this paper heavily  relies on graph theory,
especially on domatic partition problems \cite{Haynes} and 
subgraph domatic partition problems \cite{TI2015}. Notation and several fundamental 
facts required throughout the paper 
will be introduced in this section.

\subsection{Notation}

Let $G = (V,E)$ be an undirected graph, where 
the sets $V$ and $E$ represent the sets of vertices and edges, respectively.
For a node $v \in V$, the degree of $v$ is denoted by $d(v)$.
The  symbols $\delta(G)$ and $\Delta(G)$ 
represent the minimum and maximum degrees of $G$, respectively.
The {\em edge density} of $G$, denoted by $\epsilon(G)$,  is defined as
$
\epsilon(G) \defeq {|E|}/{|V|}.
$
The set of consecutive integers from $a$ to $b$ is denoted by $[a,b]$.
The symbol $\Bbb Z$ represents the set of integers.

\subsection{Subgraph domatic partition (SubDP) problems}

The directed graph version of subgraph domatic partition (SubDP) problem 
was first discussed by Wadayama, Izumi and Ono \cite{TI2015}.

In the following analysis,  we use the undirected subDP problem as a key tool. 
To present its definition, we need to clarify the definitions of 
dominating sets and domatic partitions.
\begin{definition}[Dominating set]
A dominating set $D$ of $G=(V,E)$ is a subset of $V$ such that 
any node $v \in V$ belongs to $D$ or is adjacent to a node in $D$.
\end{definition}
\begin{definition}[Domatic partition]
Let $D_1, D_2,\ldots, D_k$ be a partition of $V$; namely, $\bigcap_{i \in [1,k]} D_{i} = V$ and
any pair of subsets $D_i$ and $D_j$ is disjoint.  The partition is called a {\em domatic partition} if
all the subsets $D_1, D_2,\ldots, D_k$ are dominating sets.
\end{definition}
The domatic number $D(G)$ is the largest number of subsets in a domatic partition of $G$; i.e., 
\begin{equation}
D(G) \defeq \max\{k | D_1,D_2,\ldots, D_k \mbox{ is a domatic partition}  \}.
\end{equation}
A number of theoretical studies on domatic partitions and its applications 
have been published \cite{Haynes}. 
It is known that computing the domatic number $D(G)$ is an NP-hard problem.
The domatic number can be upper bounded by 
\begin{equation}
D(G) \le \delta(G)+1,
\end{equation}
which is called the {\em degree bound} \cite{Haynes}. 
A non-trivial lower bound proved by Feige et al. \cite{Feige} has the form:
\begin{equation}
D(G) \ge (1- o(1))(\delta(G)+1)/\ln \Delta(G),
\end{equation}
that is derived using  Lov\'asz local lemma \cite{alon}.

The SubDP problem proposed in \cite{TI2015} is a natural extension of the domatic partition problem, 
which admits choosing  an appropriate subgraph to increase the domatic number.
The solid definition of an undirected graph version of the SubDP problem is given as follows.
\begin{definition}[SubDP problem]
Let $G = (V,E)$ be a given undirected graph.
The problem to find 
the SubDP number 
$
S(G) \defeq \max_{G' \subseteq G} D(G') 
$
is called the {\em SubDP} problem.
The notation $G' \subseteq G$ means that $G'$ is a subgraph of $G$.
\end{definition}
In a broad sense, we want to have not only the SubDP number but also 
the corresponding subgraph $G'$ and the maximal domatic partition of $G'$.
It should be noted that the subDP problem is proved to be NP-hard \cite{TI2015}.

\section{$(p,q)$-transition free word sequences}

\subsection{$(p,q)$-transition free word sequences}

Let $p$ and $q$ be binary ({\tt 0} or {\tt 1}) sequences of length $k$; 
e.g., $p = {\tt 101}$ and $q = {\tt 010}$ $(k=3)$. 
The pair of sequences $p$  and $q$ is called a {\em forbidden transition pair}.
In what follows, a word  means a binary sequence of length $n (n > k)$ 
that corresponds to the set of signals exchanged in an on-chip bus.
Two binary sequences $x$ and $y$ of finite or infinite length are said to be $(p,q)$-{\em violating} 
if there is an index $i \in \Bbb Z$ satisfying 
\[
p = x_{i+1} x_{i+2} \cdots  x_{i+k},\   q =  y_{i+1} y_{i+2} \cdots   y_{i+k} 
\]
or 
\[
q = x_{i+1} x_{i+2} \cdots  x_{i+k},\   p =  y_{i+1} y_{i+2} \cdots  y_{i+k},
\]
where $x_i$ and $y_i$ denote $i$-th elements of the sequences $x$ and $y$, respectively.
Otherwise, the pair $x$ and $y$ is said to be $(p,q)$-{\em transition free}.

Our goal is to design an encoder and a decoder that generate word sequences 
(i.e., word streams exchanged in the buses) with the $(p,q)$-transition free property.
\begin{definition}[$(p,q)$-transition free word sequence]
Suppose that we have an infinite sequence of words 
$
(\ldots, a^{i}, \ldots) 
$
where $a^{i} (i \in \Bbb Z)$ is a word of length $n$. If $a^i$ and $a^{i+1}$   are $(p,q)$-transition free for any $i  \in \Bbb Z$,
then the sequence is said to be $(p,q)$-transition free word sequence.
\end{definition}

In the scenario of the data transmission over on-chip buses, 
it is reasonable to assume $(p = {\tt 10}, q= {\tt 01})$ (FTC)  or 
$(p = {\tt 101}, q = {\tt 010})$(FOC)  \cite{NoC} \cite{Pande} \cite{SS2007}.
In order to avoid (or weaken) the effect of capacitive crosstalk,  
$(p,q)$-violating  two words should not be sent consecutively.
This means that 
a word sequence sent to the buses should be a $(p,q)$-transition free word sequence. 

\subsection{Asymptotic rate of $(p,q)$-transition free word sequences}

In this paper, we will  discuss state dependent encoders 
for converting a message sequence to a  $(p,q)$-transition free word sequence.
An encoder consists of an {\em encoding function}  
${\cal E}: [1, M] \times \{{\tt 0,1}\}^n \rightarrow \{{\tt 0,1}\}^n$ 
that computes  the next state of the encoder
from a message in the range $[1,M]$ entered into the encoder  
and the current state kept in the encoder.
The symbol $M$ represents the cardinality of the message alphabet.
A set of states of the encoder consists of words in $(p,q)$-transition free word sequences.
An infinite sequence of states 
$
(\ldots, s_i, \ldots)
$
generated by the recursion $s_{i+1}= {\cal E}(m_i, s_i)$ must be a $(p,q)$-transition free word sequence for any 
message sequence $(\ldots, m_i, \ldots)$. The state $s_{i+1}$ computed by the encoding function ${\cal E}$
is sent to the channel and then the encoder state is updated to $s_{i+1}$.

A {\em decoding function} ${\cal D}: \{{\tt 0,1}\}^n \rightarrow [1,M]$ must satisfy the following consistency condition:
\begin{equation} \label{consistency}
 m = {\cal D}({\cal E}(m, s))
\end{equation}
for any $m \in [1,m]$ and any $s \in \{{\tt 0,1}\}^n$. This means that the decoding function satisfying this 
consistency condition can immediately obtain the correct message by observing an output from the encoder.
Note that the decoder does not have internal memories to keep its state, which is a desirable 
feature for a decoder to prevent error propagation due to channel noises. 

For a given forbidden transition pair $(p,q)$,  a pair $(n,M)$ is said to be {\em achievable} if
there exists a pair of encoding and decoding functions satisfying the consistency condition (\ref{consistency}).
The maximum rate of $(p,q)$-transition free word sequences is naturally defined by 
\begin{equation}
R(p,q, n) \defeq \max_{(n,M)  \mbox{ is achievable}} \frac{\log_2 M}{n}.
\end{equation}
Based on this maximum rate,
we define the asymptotic rate of  $(p,q)$-transition free word sequences by
\begin{equation}
R(p,q) \defeq \limsup_{n \rightarrow \infty} R(p,q,n).
\end{equation}
The problem setup is slightly different from the conventional problem setups of coding for constraint sequences 
that allow an encoder to have multiple words in its memories instead of only a single word assumed in this paper.
In our scenario, the number of possible words are exponential to $n$.
It thus may be reasonable to investigate the simplest encoder that requires the least hardware complexity.

\subsection{$(p,q)$-transition free graph}

It will be convenient to name 
the state transition graphs representing the $(p,q)$-transition free constraints.

\begin{figure}
\begin{center}
\includegraphics[scale=0.3]{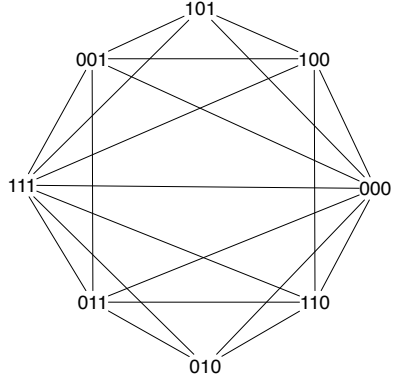} 
\caption{$({\tt 10, 01})$-transition free graph $G({\tt 10, 01}, 3)$: no $(p,q)$-violating words containing 
forbidden transition pairs ${\tt 10, 01}$ at the same bit positions,  such as ${\tt 101}$ and  ${\tt 010}$,  
are connected. }
\label{stg}
\end{center}
\end{figure}

\begin{definition}[$(p,q)$-transition free graph]
Assume a directed graph $G=(V,E)$ with $|V| = 2^n$ nodes.
Let $b$ be a bijection  between 
$V$  and all binary words of length $n$, i.e., $\{{\tt 0,1}\}^n$. 
The word corresponding to a node $v \in V$ is denoted by 
$b(v) \in \{{\tt 0,1}\}^n$.  Any two nodes $v, w \in V$ are connected if and only if $b(v)$ and $b(w)$ 
are $(p,q)$-transition free.
Then, the graph $G$ is said to be a $(p,q)$-transition free graph.
\end{definition}
By the definition, the $(p,q)$-transition free graph is uniquely determined by $(p,q)$ and $n$, which is denote by $G(p,q, n)$.
As an example, Fig.\ref{stg} presents the $({\tt 10,01})$-transition free graph $G({\tt 10, 01}, 3)$.
It can be observed that no $(p,q)$-violating two words are connected;
every pair of adjacent nodes contain no forbidden transition pair ${\tt 01} \leftrightarrow {\tt 10}$.

\section{Asymptotic growth rate of SubDP number}

In this section, we will discuss the asymptotic growth rate of SubDP number 
that has a close relationship to the maximum rate $R(p,q, n)$.

\subsection{Maximum rate and SubDP number}

Assume that a graph $G^* = (V^*, E^*)$  is the optimal subgraph of $G(p,q,n)$  that 
gives the SubDP number of $G(p,q,n)$.
There is a domatic partition of $G^*$ producing disjoint subsets of $V^*$,   $D_1, D_2, \ldots, D_{S(G(p,q,n))}$
where any subset $D_i$ is a dominating set of $G^*$.
From this partition, we can define a decoding function ${\cal D}$ by
$
{\cal D}(b(x)) \defeq i \mbox{ if } x \in D_i
$
for $x$ in $V^*$. It is evident that,
for any $x \in V^*$ and for any $i \in [1,S(G(p,q,n) )]$,
the neighbor set of  $x$ (nodes adjacent to $x$ and $x$ itself) includes
at least one node belonging to $D_i$. According to the decoding function defined above, 
an encoding function is defined by
$
{\cal E}(b(x),i) \defeq b(y)
$
for $x \in V^*$ and $i \in [1, S(G(p,q,n))]$. In this equation, the node $y \in V^*$ should be 
in the neighbor set of $x$ and belong to $D_i$. It is easy to see that the pair of these encoding and decoding functions
satisfies the consistency condition (\ref{consistency}). Since $G^*$ is a subgraph of $G(p,q,n)$, an output word sequence 
from this encoder becomes a $(p,q)$-transition free word sequence. 
In this case, we have $M = S(G(p,q,n))$ and it leads to a lower bound on the maximum rate 
of $(p,q)$-transition free word sequences:
$
R(p,q,n) \ge {\log_2 S(G(p,q,n))}/{n}.
$

On the other hand, assume that we know a pair of encoding and decoding functions $({\cal E, D})$ achieving $R(p,q, n)$.
Let $G' = (V',E')$ be a  subgraph of $G(p, q, n)$ satisfying the following conditions.
The set of nodes $V'$ is the set of nodes 
satisfying 
\begin{equation}
\forall v \in V', \forall m \in [1, 2^{n R(p,q,n)}],\   b^{-1}({\cal E}(m, b(v))) \in V'
\end{equation}
and the edge set $E'$ is given by
\begin{equation}
E' = \{ (s, b^{-1}({\cal E}(m, b(s)))) \mid s \in V', m \in [1, 2^{n R(p,q,n)}] \}.
\end{equation}
Note that both $(a,b)$ and $(b,a)$ represents the identical undirected edge.
The decoding function generates a partition of $V'$ of size $2^{n R(p,q,n)}$
and it needs to be a domatic partition.
This observation leads to the inequality $S(G(p,q,n)) \ge 2^{n R(p,q, n)}$.
Combining two inequalities on $R(p,q,n)$, we immediately have the equality on the maximum rate:
\begin{equation}
R(p,q,n) = \frac{\log_2 S(G(p,q,n))}{n}.
\end{equation}
Therefore, studying asymptotic rate of the $(p,q)$-transition free word sequences is equivalent to 
study the asymptotic behavior of the SubDP number of the $(p,q)$-transition free graph.

\subsection{Bounds on asymptotic growth rate of SubDP number}

The next theorem presents upper and lower bounds on the asymptotic growth rate of the SubDP number
for general graph sequences.

\begin{theorem} \label{agr}
Assume that a sequence of undirected graphs $G_n=(V_n, E_n) (n=1,2,3,\ldots)$
with $2^n$-nodes have a non-vanishing limit of the asymptotic growth rate of the edge density:
\[
\alpha \defeq  \lim_{n \rightarrow \infty} \frac 1 n \log_2 \epsilon(G_n) > 0.
\]
The asymptotic growth rate of the SubDP number $S(G_n)$ of this graph sequence 
is bounded as 
\begin{equation}
\alpha \le \limsup_{n \rightarrow \infty} \frac 1 n \log_2 S(G_n)  \le \frac{1+ \alpha}{2}.
\end{equation}
\end{theorem}

\noindent
(Proof of Theorem \ref{agr})
From the definition of the edge density, we have
\begin{equation}
\epsilon(G_n) = {|E_n|}/{|V_n|} = {|E_n|}/{2^n}.
\end{equation}
By exploiting a graph pruning method presented in \cite{Dielstel},
we can retrieve an induced subgraph $\tilde G \subseteq G_n$ satisfying 
\begin{equation}\label{delep}
\delta(\tilde G) \ge \epsilon(G_n) = {|E_n|}/{2^n}.
\end{equation}
The lemma due to Feige et al. \cite{Feige} guarantees the existence of
a domatic partition of $\tilde G$ with the domatic number satisfying 
\begin{eqnarray}
D(\tilde G) 
&\ge& (1- o(1))(\delta(\tilde G)+1)/\ln \Delta(\tilde G) \\
&\ge&(1- o(1))(\epsilon(G_n)+1)/\ln \Delta(G_n).
\end{eqnarray}
In the derivation of the last inequality, the inequality (\ref{delep}) was used.
Due to this inequality,  we can derive a lower bound on the asymptotic growth rate 
of $S(\tilde G)$ in the following way:
\begin{eqnarray}
\limsup_{n \rightarrow \infty} \frac 1 n \log_2 S(G) 
&\ge&
\limsup_{n \rightarrow \infty}\frac 1 n \log_2 D(\tilde G) \\
&\ge& \lim_{n \rightarrow \infty}\frac 1 n \log_2 \epsilon(G_n) = \alpha.
\end{eqnarray}
We then consider the opposite direction.
Let $G^* =  (V^*, E^*)$ be the subgraph of $G_n$ that gives the SubDP number $S(G_n)$.
For any node $v \in G^*$, the degree of $v$ must satisfy 
$
d(v) \ge S(G_n) -1.
$
By using this inequality on $d(v)$, we have a sequence of inequalities:
\begin{eqnarray}
|E_n| \ge |E^*| &=& (1/2) \sum_{v \in V^*} d(v) \\
&\ge& (1/2) |V^*| (S(G_n) -1) \\
&\ge& (1/2)  S(G_n) (S(G_n) -1).
\end{eqnarray}
The last inequality is based on a simple fact that $|V^*| \ge S(G_n)$.
In summary, the quantity $2 |V_n|\epsilon(G_n)$  can be lower bounded by 
$
2 |V_n|\epsilon(G_n)  = 2|E_n| \ge S(G_n) (S(G_n) -1).
$
Taking limsup on the both sides of the above inequality, we immediately obtain
\begin{equation}
\lim_{n \rightarrow \infty} \frac 1 n \log_2 2|V_n| \epsilon(G_n)\ge \limsup_{n \rightarrow \infty} 
\frac 1 n \log_2 S(G_n) (S(G_n) -1)
\end{equation}
that can be reduced to the upper bound
\begin{equation}
\limsup_{n \rightarrow \infty} \frac 1 n \log_2 S(G_n)  \le \frac{1+ \alpha}{2}
\end{equation}
in the claim of the theorem. \hfill\qed

Note that the bounds shown in Theorem \ref{agr} are sharp.
Suppose that the graph $G_n$ is the complete graph of size $2^n$.
In this case, we have
\begin{equation}
\alpha = \lim_{n \rightarrow \infty} \frac 1 n \log_2 \frac{2^n(2^n-1)}{2^{n+1}} = 1.
\end{equation}
Substituting $\alpha=1$ into the bounds, the lower bound coincides with the upper bounds
and we obtain
$
\limsup_{n \rightarrow \infty} (1/n) \log_2 S(G_n) = 1.
$

\section{Asymptotic growth rate of edge density}

In this section, we will describe a method for evaluating the number of edges
of the $(p,q)$-transition free graph that is required for deriving the asymptotic growth rate of
the edge density:
$
\alpha(p,q)  \defeq \lim_{n \rightarrow \infty} (1/n) \log_2  \epsilon(G(p,q,n)).
$

\subsection{Size of edge set of $G(p,q,n)$}
Let $N(p,q,n)$ be the number of the $(p,q)$-transition free word pairs; i.e.,  
\begin{equation}
N(p,q,n) \defeq |\{(a,b) \in \{{\tt 0,1} \}^n \times \{{\tt 0,1}\}^n \mid a, b \mbox{: $(p,q)$-tr. free} \}|.
\end{equation}
The number $N(p,q, n)$ can be used for counting the size of 
the edge set, denoted by $E(p,q,n)$ of $G(p,q,n)$:
\begin{equation}
|E(p,q, n)| = \frac{N(p,q, n) - 2^n}{2}.
\end{equation}
The term $2^n$ in the numerator is included to exclude the pairs consisting of the same word.
The denominator compensates over counts on edges; 
$(a,b)$ and $(b,a)$ represents an identical edge in $G(p,q,n)$.

\subsection{Counting of $({\tt 10,01})$-transition free word pairs}

This subsection describes how to count $N(p,q, n)$.
In order to simplify the discussion, we will focus on the simplest case (FTC) where $p={\tt 10}, q={\tt 01} $ 
in this subsection (general cases are to be discussed later). 

\begin{figure}
\begin{center}
\includegraphics[scale=0.3]{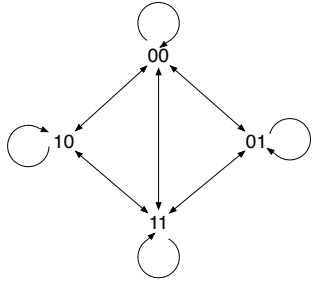} 
\caption{State transition graph for $({\tt 10, 01})$-transition free sequence pair: the label sequence of any walk in this graph 
corresponds to a $({\tt 10, 01})$-transition free sequence pair.}
\label{transitive}
\end{center}
\end{figure}

Suppose the situation where 
an infinite sequence 
$
\ldots, (a_i,b_i), (a_{i+1}, b_{i+1}),  \ldots
$
follows the state transition graph depicted in Fig. \ref{transitive} where $(a_i,b_i) \in \{{\tt 0,1}\}^2$ for $i \in \Bbb Z$.
Since there are no state transitions between ${\tt  (1,0)} \leftrightarrow {\tt (1,0)}$ in the state transition graph,
the two sequences
$
(\ldots, a_i, \ldots)
$
and 
$
(\ldots, b_i, \ldots)
$
are $({\tt 10},{\tt 01})$-transition free. Furthermore, any $({\tt 10},{\tt 01})$-transition free sequence pair
corresponds to an allowable walk in the state transition graph.
Thus, calculation of  $N({\tt 10},{\tt 01},n)$ can be done by counting the number 
of allowable walks of length $n$ in the state transition graph.
It is simply carried out by using matrix multiplication:
\begin{equation}
N({\tt 10},{\tt 01},n) = 
(1 1 1  1) A^{n-1}
\left(
\begin{array}{c}
1  \\
1  \\
1  \\
1  \\
\end{array}
\right),\ 
A \defeq 
\left(
\begin{array}{cccc}
1 & 1 & 1 & 1 \\
1 & 1 & 0 & 1 \\
1 & 0 & 1 & 1 \\
1 & 1 & 1 & 1 \\
\end{array}
\right),
\end{equation}
where $A$ is an adjacent matrix of the state transition graph in Fig. \ref{transitive}.
The largest eigenvalue of $A$ is 
\[
\lambda_{max} = \frac 1 2 \left(3 + \sqrt{17} \right)
\]
and the corresponding eigenvector is
\begin{equation}
p_{max} = \left(1, - 1 + \frac 1 4 \left(3 + \sqrt{17} \right), - 1 + \frac 1 4 \left(3 + \sqrt{17} \right), 1 \right)^T.
\end{equation}

It is well known that, for any nonzero initial vector $x$, 
$A^n x$ approaches to $\beta \lambda_{max}^n p_{max}$ when $n$ goes to infinity where 
$\beta$ is a real constant. By using this fact, we immediately have 
\begin{equation}
\lim_{n \rightarrow \infty} \frac 1 n \log_2 N({\tt 10},{\tt 01},n)  
= \lambda_{max} = \frac 1 2 \left(3 + \sqrt{17} \right). 
\end{equation}
We are now ready to evaluate $\alpha({\tt 10},{\tt 01})$:
\begin{eqnarray}
\alpha({\tt 10},{\tt 01}) &=& \lim_{n \rightarrow \infty} \frac 1 n \log_2  \epsilon(G({\tt 10},{\tt 01},n)) \\
&=& \lim_{n \rightarrow \infty} \frac 1 n \log_2  \left(\frac{|E({\tt 10},{\tt 01},n) |}{2^n} \right) \\
&=& \lim_{n \rightarrow \infty} \frac 1 n \log_2  \left(\frac{\lambda_{max}^n (1 + o(1)) }{2^n} \right) \\
&=& \log_2 \left({\lambda_{max}}/ {2} \right) \\
&=& -2 + \log_2 \left(3 + \sqrt{17}  \right) \simeq  0.8325.
\end{eqnarray}

Substituting the value $\alpha({\tt 10},{\tt 01})$ into the upper and lower bounds 
in Theorem \ref{agr},  the following corollary is obtained.
\begin{corollary} \label{arate}
The asymptotic rate $R({\tt 10},{\tt 01})$ for the $({\tt 10}, {\tt 01})$-transition free word sequences is bounded as
\begin{equation}
-2 + \log_2 \left(3 + \sqrt{17}  \right) \le R({\tt 10},{\tt 01}) \le \frac{-1 + \log_2 \left(3 + \sqrt{17}  \right)}{2}.
\end{equation}
\end{corollary}
Note that the values in the bounds can be approximated as $0.8325 \le R({\tt 10},{\tt 01}) \le 0.9162$.

It is shown in \cite{Victor} that 
the asymptotic growth rate of the minimum degree of the $({\tt 10},{\tt 01})$-transition free graphs
is given by
\begin{equation}
\lim_{n \rightarrow \infty} \frac 1 n \log_2 \delta(G({\tt 10},{\tt 01},n))  = \log_2 \left( \frac{1+\sqrt{5}}{2} \right) \simeq 0.6942.
\end{equation}
This means that the asymptotic rate of coding schemes
constructed directly from the domatic partition of $G({\tt 10},{\tt 01},n)$ cannot exceed 0.6942 
because 
the domatic number is less than or equal to $\delta(G({\tt 10},{\tt 01},n))+1$.
On the other hand, Corollary \ref{arate} gives a guarantee of existence of coding schemes
with the asymptotic rate beyond 0.8325.
An apparent implication of this observation is that finding an appropriate subset in $G(p,q,n)$
is crucial for achieving near optima rate when $n$ is sufficiently large.
In other words, considering SubDP problems on $G(p,q,n)$ is indispensable to design efficient 
long codes for the $(p,q)$-transition free word sequences.
Note that Victor and Keutzer \cite{Victor} reported that 
the asymptotic rate of stateless coding for the $({\tt 10},{\tt 01})$-transition free word sequences 
cannot exceed 0.6942.

\subsection{Counting for general $(p,q)$-transition free sequence pairs}

The key of successful calculation of the number of edges in $G({\tt 10,01},n)$
was to define an appropriate state transition graph representing  all the 
$({\tt 10,01})$-transition free sequence pairs. The same technique can be extended to general cases.
Assume that a directed graph ${\cal G}$ with $2^{2k-2}$ nodes is given
and that each node is labeled with a binary $2k-2$ tuple (i.e., there is a bijection between 
the node set and $\{{\tt 0,1}\}^{2k-2}$).
If and only if, for any pair of $(p,q)$-transition free sequences $a=(\ldots, a_i, \ldots ), b=(\ldots, b_i, \ldots )$ and 
for any index $i \in \Bbb Z$, 
an edge from the node with label
\[
(a_{i+1}, b_{i+1},a_{i+2}, b_{i+2}, \ldots, a_{i+k-1}, b_{i+k-1}) \in \{{\tt 0,1}\}^{2k-2}
\]
to the node with label
\[
(a_{i+2}, b_{i+2},a_{i+3}, b_{i+3}, \ldots, a_{i+k}, b_{i+k}) \in \{{\tt 0,1}\}^{2k-2}
\]
exists, then the graph ${\cal G}$ is said to be the $(p,q)$-{\em transition free pair graph}.
Any $(p,q)$-transition free sequence pair corresponds to a walk on the $(p,q)$-transition free pair graph.
This means that counting $N(p,q,n)$ is equivalent to count the number of possible walks of length $n$ 
in the $(p,q)$-transition free pair graph. As in the case of the previous subsection, 
we can use the same technique to evaluate the growth rate of $N(p,q,n)$.
The largest eigenvalue of the adjacent matrix of $(p,q)$-transition free pair graph dominates 
the asymptotic behavior of the number of the edge set of $G(p,q,n)$.
For example, the $(p={\tt 101}, q = {\tt 010})$-transition free pair graph $(k=3)$ consists of 16-nodes.
Except for the two nodes corresponding to the forbidden transition pair,  every nodes in the graph has outbound degree 4.
Precisely speaking, the edges ${\tt 0110 \rightarrow 1001}$ and ${\tt 1001 \rightarrow 0110}$  are missing.
From the maximum eigenvalue of the adjacent matrix of this graph, we can immediately evaluate the asymptotic growth rate of
the edge density as
$
\alpha({\tt 101,010}) \simeq 0.9636.
$


\begin{thebibliography}{99}

\bibitem{alon}
N. Alon and J.H. Spencer,
``The probabilistic method, '' 2nd ed., John Wiley \& Sons,  2000.


\bibitem{Feige}
U. Feige, M. M. Halld\'orsson, Gu. Kortsarz, and A. Srinivasan,
``Approximating the domatic number, ''
SIAM J. Comput., pp.172--195, vol.32,  2002.

\bibitem{Dielstel}
R. Diestel,
``Graph theory,''
Springer-Verlag, 2006.


\bibitem{Haynes}
T. W. Haynes, S. T. Hedetniemi, and P. J. Slater,
``Fundamentals of domination in graphs,''
Marcel Dekker, 1998.

%


\bibitem{TI2015}
T. Wadayama, T. Izumi, and H.Ono,
``Subgraph domatic problem and writing capacity of memory devices with restricted state transitions, ''
Proceedings of IEEE International Symposium on Information Theory (ISIT),
pp.1307--1311, June, HongKong, 2015.

\bibitem{NoC}
S. Kundu and S. Chattopadhyay,
``Network-on-chip, the next generation of system-on-chip integration, ''
CRC Press, 2015.

\bibitem{Pande}
P.P. Pande, A. Ganguly, B. Reero, B. Belzer, and C. Grecu,
``Design of low power \& reliable networks on chip through joint crosstalk avoidance and forward 
error correction coding,''  Proceedings of IEEE International Symposium on Defect and Fault-Tolerance in VLSI
Systems, Arlington, pp. 466--476, 2006.


\bibitem{SS2007}
S.R. Sridhara and N.R. Shanbhag,
``Coding for reliable on-chip buses: a class of fundamental bounds and practical codes,''
IEEE Transactions on CAD of Integrated Circuits and Systems, vol.26, no.5, pp.977--982, 2007.


\bibitem{Victor}
B. Victor and K. Keutzer,
``Bus encoding to prevent crosstalk delay, ''
 Proceedings of IEEE/ACM ICCAD, pp.57-63, Nov. 2001.

\bibitem{Niesen}
U.Niesen and S.Kudekar,
``Joint crosstalk-avoidance and error-correction coding for parallel data buses, ''
arXiv:1601.04961, 2016.

\end{thebibliography}
\end{document}